%% file: main_regular.tex
\documentclass{article}
\input{packages.tex}
\input{macro.tex}
\input{acronyms.tex}
\ninept
\title{Discriminative Neural Clustering for Speaker Diarisation}%
\name{Qiujia Li$^\star$, Florian L. Kreyssig$^\star$, Chao Zhang \& Philip C. Woodland\thanks{$^\star$Indicates first authors. Ordering determined by coin flip. Kreyssig is funded by a EPSRC DTP. Li is funded by a Peterhouse Graduate Studentship.}\vspace{-0.5em}}
\address{Cambridge University Engineering Dept., Trumpington St., Cambridge, CB2 1PZ U.K.\\
\small{\texttt{\{ql264,flk24,cz277,pcw\}@eng.cam.ac.uk}}\vspace{-0.5em}}
\copyrightnotice{\copyright IEEE 2020}
\begin{document}
%
\maketitle

\input{text/abs.tex}
\input{text/intro.tex}
\input{text/neural_clustering.tex}
\input{text/data_aug.tex}
\input{text/setup.tex}

\input{text/curlurn.tex}
\input{text/conclusion.tex}

\section{References}
\begingroup
\renewcommand{\section}[2]{}
\bibliographystyle{IEEEbib}
\bibliography{refs}
\endgroup

\end{document}

%% file: packages.tex
\usepackage[preprint]{spconf}

\usepackage{color,xcolor}
\usepackage{epsfig}
\usepackage{graphicx}

\usepackage{array}
\usepackage{booktabs}
\usepackage{colortbl}
\usepackage{multirow}
\usepackage{float}
\usepackage[labelformat=simple]{subcaption}

\usepackage{footnote}
\makesavenoteenv{tabular}
\makesavenoteenv{table}

\usepackage{amsmath,amsfonts,amssymb,bm}
\usepackage[super]{nth}

\usepackage{changepage}
\usepackage{extramarks}
\usepackage{fancyhdr}
\usepackage{lastpage}
\usepackage{setspace}
\usepackage{soul}
\usepackage{xspace}

\usepackage{url}
\usepackage{hyperref}
\hypersetup{colorlinks=True, urlcolor=black}
\usepackage[numbers,sort&compress]{natbib}

\usepackage{algorithm, algorithmic}
\usepackage{enumitem}
\usepackage{verbatim}
\usepackage{pifont}
\usepackage[acronyms]{glossaries}
\glsdisablehyper
\usepackage{caption}
\usepackage{subcaption}

%% file: macro.tex
\newcommand{\sect}[1]{Section~\ref{sec:#1}}
\newcommand{\sectdot}[1]{Sec.~\ref{sec:#1}}
\newcommand{\ssect}[1]{Section~\ref{ssec:#1}}
\newcommand{\ssectdot}[1]{Sec.~\ref{ssec:#1}}

\newcommand{\eqndot}[1]{Eqn.~(\ref{eqn:#1})}
\newcommand{\fig}[1]{Figure~\ref{fig:#1}}
\newcommand{\figdot}[1]{Fig.~\ref{fig:#1}}
\newcommand{\tbl}[1]{Table~\ref{tab:#1}}

\newcommand{\twosect}[2]{Sections~\ref{sec:#1} and \ref{sec:#2}}
\newcommand{\twosectdot}[2]{Sec.~\ref{sec:#1} and \ref{sec:#2}}

\newcommand{\ignore}[1]{}

\makeatletter
\DeclareRobustCommand\onedot{\futurelet\@let@token\@onedot}
\def\@onedot{\ifx\@let@token.\else.\null\fi\xspace}

\def\eg{\emph{e.g}\onedot} 
\def\ie{\emph{i.e}\onedot} 
 
 \def\vs{\emph{vs}\onedot}
 
\def\etal{\emph{et al}\onedot}

\makeatother

\definecolor{MyDarkBlue}{rgb}{0,0.08,1}
\definecolor{MyDarkGreen}{rgb}{0.02,0.6,0.02}
\definecolor{MyDarkRed}{rgb}{0.8,0.02,0.02}
\definecolor{MyDarkOrange}{rgb}{0.40,0.2,0.02}
\definecolor{MyPurple}{RGB}{111,0,255}
\definecolor{MyRed}{rgb}{1.0,0.0,0.0}
\definecolor{MyGold}{rgb}{0.75,0.6,0.12}
\definecolor{MyDarkgray}{rgb}{0.66, 0.66, 0.66}

\def\presec{\vspace{-0em}}
\def\postsec{\vspace{-0em}}

\definecolor{spkA}{HTML}{1f77b4}
\definecolor{spkB}{HTML}{ff7f0e}
\definecolor{spkC}{HTML}{2ca02c}
\definecolor{spkD}{HTML}{d62728}
\definecolor{spkE}{HTML}{9467bd}
\newcommand{\spkA}{\textcolor{spkA}{A\;}}
\newcommand{\spkB}{\textcolor{spkB}{B\;}}
\newcommand{\spkC}{\textcolor{spkC}{C\;}}
\newcommand{\spkD}{\textcolor{spkD}{D\;}}
\newcommand{\spkE}{\textcolor{spkE}{E\;}}

\newcommand{\meetinga}{{\ensuremath{\mathtt{meeting}}}\xspace}
\newcommand{\globala}{{\ensuremath{\mathtt{global}}}\xspace}
\newcommand{\nonea}{{\ensuremath{\mathtt{none}}}\xspace}
\newcommand\mydots{\makebox[1em][c]{.\hfil.\hfil.}}

%% file: acronyms.tex
\newacronym{rnn}{RNN}{recurrent neural network}
\newacronym{ann}{ANN}{artificial neural network}
\newacronym{mlp}{MLP}{Multi-layer perceptron}
\newacronym{asr}{ASR}{automatic speech recognition}
\newacronym{tdnn}{TDNN}{time-delay neural network}
\newacronym{fc}{FC}{fully-connected}
\newacronym{asoft}{ASoftmax}{angular softmax}
\newacronym{dnc}{DNC}{Discriminative Neural Clustering}
\newacronym{mha}{MHA}{multi-head attention}
\newacronym{vad}{VAD}{voice activity detection}
\newacronym{cpd}{CPD}{speaker change point detection}
\newacronym{ser}{SER}{speaker error rate}
\newacronym{der}{DER}{diarisation error rate}
\newacronym{uisrnn}{UIS-RNN}{unbounded interleaved-state \gls{rnn}}
\newacronym{std}{std}{standard deviation}
\newacronym{tsne}{t-SNE}{t-distributed stochastic neighbour embedding}
\newacronym{sc}{SC}{spectral clustering}
\newacronym{mdm}{MDM}{multiple distance microphone}
\newacronym{daug}{Diac-Aug}{Diaconis augmentation}
\newacronym{cl}{CL}{curriculum learning}
\newacronym{pit}{PIT}{permutation invariant training}

%% file: text/abs.tex
\begin{abstract}
In this paper, we propose \gls{dnc} that formulates data clustering with a maximum number of clusters as a supervised sequence-to-sequence learning problem. Compared to traditional unsupervised clustering algorithms, \gls{dnc} learns clustering patterns from training data without requiring
an explicit definition of a similarity measure.
An implementation of \gls{dnc} based on the Transformer architecture is shown to be effective on a speaker diarisation task using the challenging AMI dataset. 
Since AMI contains only 147 complete meetings as individual input sequences, data scarcity is a significant issue for training a Transformer model for \gls{dnc}. 
Accordingly, this paper proposes three data augmentation schemes: sub-sequence randomisation, input vector randomisation, and Diaconis augmentation, which generates new data samples by rotating the entire input sequence of $L_2$-normalised speaker embeddings.
Experimental results on AMI show that \gls{dnc} achieves a reduction in \gls{ser} of 29.4\% relative to spectral clustering.
\end{abstract}

\begin{keywords}
speaker diarisation, supervised clustering, discriminative neural clustering, Transformer
\end{keywords}


%% file: text/intro.tex
\glsreset{dnc}
\glsreset{daug}
\glsreset{ser}
\presec
\section{Introduction}
\postsec
\label{sec:intro}
Clustering is the task of grouping data samples into multiple clusters such that a sample of a particular cluster is more similar to samples of that cluster than to those of others clusters. The performance of clustering algorithms is crucial in many applications.
For speaker diarisation, the task to determine ``who spoke when'' in a multi-talker audio stream~\cite{tranter2006overview,Anguera2012SpeakerDA}, such as a meeting or a conversation, the audio stream is usually first divided into many speaker-homogeneous segments. These segments are then clustered into groups that correspond to their speaker identities.
The input samples to the clustering algorithms used for speaker diarisation are often speaker embeddings~\cite{dehak2011ivec,snyder2017deep,snyder2018xvec,wan2018triplet}, \ie fixed dimensional vectors that represent the speaker identity of each segment in a distributed fashion.
Commonly used clustering algorithms, such as agglomerative clustering~\cite{karanasou2015mgbdiar,garcia2017dvecdiar,sell2018diarization}, K-means clustering~\cite{shum2013unsupdiar,dimitriadis2017developing} and spectral clustering~\cite{ning2006spectral,wang2018diarLSTM,sun2019diarselfatt}, 
are mostly unsupervised and model-free, often leveraging pre-defined distance measures and hyper-parameters to determine the similarity between data samples.

Clustering is challenging and inherently ambiguous when samples of different clusters are not well separated in the feature space. Various loss functions~\cite{wan2018triplet,deepclustering,le2018robust} and model structures~\cite{sun2019diarselfatt,lin2019lstm} have been designed to extract speaker embeddings that are better suited for unsupervised clustering in speaker diarisation pipelines. These methods often try to match some assumptions made by the clustering algorithms that are related to the underlying data distributions or distance measures. From this perspective, using a parametric model to learn how to cluster the embeddings from the training data is more desirable since it has the potential to avoid enforcing such assumptions. Furthermore, algorithms such as K-means and spectral clustering require an iterative process with multiple sets of randomly initialised values, and the related hyper-parameters often need to be carefully adjusted. 
Although supervised learning can be a desirable solution to the clustering problem when training examples are available, it is often difficult to use existing parametric classifiers for clustering purposes directly.
For example, speaker diarisation requires speech segments to be assigned to speakers with unknown identities. Hence, a clustering algorithm should not be associating a target with a particular speaker. Rather than determining the absolute speaker identity of each segment as in a speaker classification task, it is the relative speaker identities across all segments that are of interest.

In this paper, we propose \gls{dnc}, which uses a sequence-to-sequence neural network, in particular a Transformer model~\cite{Vaswani2017AttentionIA}, originally designed for sequence classification to perform supervised clustering for speaker diarisation.
In contrast to end-to-end speech recognition using sequence-to-sequence  models~\cite{Bahdanau2016EndtoendAL,Chan2016ListenAA}, where each input sequence is an utterance with a duration of tens of seconds, speaker diarisation requires each input sequence to be the audio stream of a complete conversation or a meeting with a much longer duration of tens of minutes or even hours. To avoid dealing with overly-long sequences, each feature vector in the input sequences of \gls{dnc} corresponds to the speaker embedding of a speaker-homogeneous segment rather than a frame. 
Treating one input sequence as one conversation or one meeting causes the amount of supervised training data for clustering to be severely limited. For example, in the augmented multi-party interaction (AMI) dataset, widely used for speaker diarisation~\cite{carletta2005ami}, 
only 147 meetings exist that in turn can be used as individual training sequences. The three data augmentation schemes that this paper proposes to overcome the data scarcity are called sub-sequence randomisation, input vector randomisation, and the \gls{daug}.
\begin{itemize}
    \item Sub-sequence randomisation selects many sub-meetings with randomised start and end points as the augmented input sequences.
    \item Input vector randomisation generates augmented input sequences by keeping the cluster label sequence of the original training example, replacing each cluster with a randomly selected speaker identity and selecting feature vectors corresponding to the speaker identity at random.
    \item As the speaker embeddings used are $L_2$-normalised, 
    new training sequences can be generated by rotating the entire input sequence to a different region of the hypersphere. This novel scheme is termed \gls{daug}.
\end{itemize}
A specific, unambiguous ordering of the cluster labels is enforced.
By only allowing one possible label sequence, this, in addition to the augmentation schemes, helps mitigate the data scarcity issue. The augmentation schemes also enable \gls{dnc} to learn the importance of relative speaker identities (the cluster labels) rather than absolute speaker identities across segments, which allows \gls{dnc} to perform speaker clustering with a simple cross-entropy training loss function.

This paper is organised as follows. \sect{related_work} reviews the related work. \sect{neural_clustering} introduces the proposed DNC method. \sect{data_aug} describes the proposed data augmentation methods. \twosect{setup}{results} present the experimental setup and results on the AMI dataset. Finally, analysis and conclusions are given in \twosectdot{analysis}{conclusion}.

%% file: text/neural_clustering.tex
\presec
\section{Related Work}
\postsec
\label{sec:related_work}
Unsupervised clustering algorithms, such as agglomerative clustering, K-means and spectral clustering, are widely used for clustering speaker embeddings in speaker diarisation~\cite{karanasou2015mgbdiar,garcia2017dvecdiar,sell2018diarization,shum2013unsupdiar,dimitriadis2017developing,ning2006spectral,wang2018diarLSTM,sun2019diarselfatt}. They often assume the underlying data distributions to be Gaussian or measure the distance between speaker embeddings using the cosine similarity. Accordingly, speaker embeddings have been trained to minimise the cosine similarities between the embeddings of different speakers while maximising those of the same speaker. 
Compared to unsupervised clustering algorithms, \gls{dnc} does not make any assumptions about the data distribution or a distance measure. 
Recently, graph neural networks have been used to improve spectral clustering~\cite{Shaham2018SpectralNetSC,Wang2020GNNSC4Dia}.

Supervised clustering has been applied to speaker clustering by, which uses {a} \gls{rnn} as a generative model in an approach called \gls{uisrnn}~\cite{zhang2019fully}.
All speakers in a meeting share the same set of \gls{rnn} parameters but have different state sequences. The \gls{uisrnn} is capable of handling an unlimited number of speakers by assuming that the occurrence of speakers follows the same distance-dependent Chinese restaurant process (ddCRP)~\cite{zhang2019fully}.
Although it has been demonstrated that \gls{uisrnn} worked well on the CALLHOME telephone conversation dataset, the task is relatively special since two speakers dominate about 90\% of the speaking time and the average duration of a conversation is around 2 minutes.
In contrast, AMI is a more general and more difficult task since each meeting has four or more active speakers who can speak at any time and in any order and a meeting lasts for more than 30 minutes on average, which makes it not as suitable for ddCRP. 
During the generative process of \gls{uisrnn}, a speaker embedding is modelled by a normal distribution with identity covariance matrix and the mean given by the RNN. The \gls{dnc} model does not impose these assumptions but requires to know the maximum number of speakers allowed.

Recent work on end-to-end models for diarisation~\cite{fujita2019end,fujita2019endself} combines neural networks with \gls{pit}~\cite{Yu2017PermutationIT} to directly produce cluster assignments from acoustic features. However, the method assumes independence of the output labels, even though they are strongly correlated in practice. In contrast, \gls{dnc} uses the sequence-to-sequence structure that conditions each output on the full output history. Moreover, \gls{pit} requires repeating each training sample $K$-factorial times and can be very expensive for large $K$, while \gls{dnc} uses a permutation-free training loss by enforcing a specific way of ordering the output label sequence. 

\presec
\section{Discriminative Neural Clustering}
\postsec
\label{sec:neural_clustering}
Clustering groups multiple high-dimensional feature vectors according to the similarity of certain desired attributes. 
In other words, for an input sequence with $N$ feature vectors,  $\bm{X}\!=\![\bm{x}_1,\mydots,\bm{x}_N]^{\text T}$, a clustering algorithm assigns each feature vector $\bm{x}_i$ a cluster label. 
As long as each cluster is associated with a unique identity label, permutating the cluster labels should not affect the clustering outcome.
For example, for two clusters, interchanging the cluster labels does not affect the clustering outcome.
Therefore, assuming the maximum number of clusters is known, the task of clustering can be considered as a special sequence-to-sequence classification problem using the attention-based encoder-decoder structure~\cite{Bahdanau2014NeuralMT}, in which the input and output sequences have equal lengths and each output target represents the cluster label of the corresponding input.

When applying \gls{dnc} to a clustering task, with input sequence is $\bm{X}$, each feature vector $\bm{x}_i$ has an underlying identity $z_i$. At inference time  \gls{dnc} attempts to assign $\bm{x}_i$ to a cluster label $y_i\!\in\!\mathbb{N}_1$, such that all feature vectors in the input sequence that belong to the same identity are assigned the same cluster label.
Therefore, instead of the absolute identities $z_i$ assigned to each $\bm{x}_i$, the relative cluster labels $y_{1:N}$ across $\bm{x}_1,\ldots,\bm{x}_N$ matters. 
In the context of speaker diarisation, $\bm{X}$ represents a meeting where each feature vector $\bm{x}_i$ is the speaker embedding extracted from a speech segment of the audio, $z_i$ is the actual speaker identity, and $y_i$ is the cluster label assigned to the corresponding segment within that meeting.
For \gls{dnc} to be applicable, multiple data samples $\left(\bm{X},z_{1:N}\right)$ have to be available for training. These are mapped to $\left(\bm{X},y_{1:N}\right)$ by enforcing a specific ordering of the cluster labels $y_{1:N}$ as shown in the following examples.
\begin{table}[!htbp]
\vspace{-0.5em}
    \centering
    \begin{tabular}{lll}
        identity sequence $z_{1:N}$ & & cluster label sequence $y_{1:N}$\\
        \midrule
        \spkE\spkA\spkC\spkA\spkE\spkE\spkC & & $\mathtt{1\;2\;3\;2\;1\;1\;3}$ \\
        \spkA\spkC\spkA\spkB\spkB\spkC\spkD\spkB\spkD & & $\mathtt{1\;2\;1\;3\;3\;2\;4\;3\;4}$
    \end{tabular}
     \vspace{-0.5em}
\end{table}

\noindent Here, \{A, B, C, D, E\} are five different speaker identities. In the first meeting, only \{A, C, E\} participate and `E' was the first person to speak, so for \gls{dnc} the cluster label `$\mathtt{1}$' is assigned to the first input and whenever she speaks again. When a new speaker speaks, `A' in this case, \gls{dnc} is trained to assign the incremented cluster label `$\mathtt{2}$' and similarly thereafter. As shown in the second example, \gls{dnc} will assign `$\mathtt{1}$' to the speaker `A' and `$\mathtt{2}$' to speaker `C' according to the order of appearance.

\presec
\subsection{Sequence-to-sequence Models for Clustering}
\postsec
\label{ssec:dnc}
In practice, cluster boundaries are rarely clear. Without prior information, making clustering decisions (deciding how many clusters and cluster boundaries) is intrinsically ambiguous. Learning domain-specific knowledge contained within the data samples can help to resolve ambiguities. 
A model attempting to determine $y_i$, the cluster assignment of $\bm{x}_i$, should condition that decision on the entire input sequence $\bm{X}$ and also on all assignments made for previous feature vectors $y_{0:i-1}$. 
Hence, we propose to model clustering with a discriminative sequence-to-sequence model:
\begin{equation}
    P(y_{1:N}|\bm{X}) = \prod_{i=1}^{N}P(y_{i}|y_{0:i-1},\bm{X}),
\end{equation}
where $y_0$ denotes a start-of-sequence token. For the model to be end-to-end trainable, a sequence-to-sequence model is used, more specifically an attention-based encoder-decoder model~\cite{Bahdanau2014NeuralMT}, which generally consists of two components
\begin{align}
    \bm{H} & = \textsc{Encoder}(\bm{X})\label{eqn:encoder}\\
    y_i    & = \textsc{Decoder}(y_{0:i-1}, \bm{H}),\label{eqn:decoder}
\end{align}
where $\bm{H}=[\bm{h}_1,\mydots,\bm{h}_N]^{\text T}$ is an encoded hidden representation sequence of $\bm{X}$. 
The encoder of \gls{dnc} transforms the input sequences such that the hidden representation captures the similarity between different feature vectors. 
As shown in \eqndot{decoder}, the decoder assigns the next cluster label according to the cluster labels assigned to all previous feature vectors. If the model believes the next feature vector to belong to one of the existing clusters, it will look up the previously assigned cluster label and reuse it. Otherwise, it will assign a new cluster label to the feature vector.

\presec
\subsection{Clustering Using Transformers}
\postsec
For both the \textsc{Encoder} and \textsc{Decoder}, 
various neural network architectures can be used.
As a particular type of \gls{dnc}, this paper uses the Transformer architecture~\cite{Vaswani2017AttentionIA}. The choice is motivated by the Transformer's ability to handle long input sequences and its superior performance across many tasks. 
A core operation of the Transformer is \gls{mha} which is based on scaled dot-product attention. For a model with $H$ heads of $D$ dimensions each, scaled dot-product attention is defined by:
\begin{equation}
    \textsc{Attention}(\Phi,\Psi) = \mathtt{softmax}\Bigg(\dfrac{\big(\bm{\Phi}\bm{W}_q\big)\big(\bm{\Psi}\bm{W}_k\big)^{\text T}}{\sqrt{D}}\Bigg)\big(\bm{\Psi}\bm{W}_v\big),
    \label{eqn:attention}
\end{equation}
where the $\bm{\Phi}\;\!\!\!\in\;\!\!\!\mathbb{R}^{L\!\times\!D}$ are called queries. The matrices $\bm{\Psi}\!\in\!\mathbb{R}^{L'\!\times\!D}$ are called keys and values, which are identical. Queries and keys are multiplied by query and key projection matrices $\{\bm{W}_q,\bm{W}_k\}\!\in\!\mathbb{R}^{HD\times D}$. The $\mathtt{softmax}$ operation is performed row-wise. Per query vector, \mbox{\textsc{Attention}} returns a weighted sum of value vectors $\bm{\Psi}$ projected by $\bm{W}_v\!\in\!\mathbb{R}^{HD\times\!D}$. \gls{mha} has multiple heads, with each of them being a dot-product attention defined in \eqndot{attention} with a distinct set of parameters. Dot-product attention can be viewed as cosine similarity given unit vectors, where a large cosine similarity corresponds to a large attention weight.  
The outputs of the multiple heads are concatenated ($\mathtt{cat}$) and then transformed by an output matrix $\bm{W}_o\!\in\!\mathbb{R}^{HD\times\!HD}$, as shown in \eqndot{mha}.
\begin{align}
    \nonumber\textsc{Head}_h(\bm{\tilde{X}})&=\begin{cases}
        \textsc{Attention}(\bm{\tilde{X}},\bm{\tilde{X}}) & \text{for self-attention}\\
        \textsc{Attention}(\bm{\tilde{X}},\bm{H})  & \text{for source-attention}
    \end{cases}\\
    \textsc{MHA}(\bm{\tilde{X}})&= \mathtt{cat}(\textsc{Head}_1(\bm{\tilde{X}}),\mydots,\textsc{head}_H(\bm{\tilde{X}}))\bm{W}_o\label{eqn:mha}
\end{align}
An encoder block consists of a \emph{self-attention} layer followed by a feed forward layer.
For the first encoder block 
$\bm{\tilde{X}}$ is the input feature sequence $\bm{X}$ and for later layers $\bm{\tilde{X}}$ is the output from the previous block.
A decoder block consists of a \emph{self-attention} layer, a \emph{source-attention} layer and a feed forward layer. For a self-attention layer in a decoder block, the input $\bm{\tilde{X}}$ is the embedding of the preceding cluster label sequence $y_{0:i-1}$ for the first block and for later layers $\bm{\tilde{X}}$ is the output of the previous block. For a source-attention layer in a decoder block, the input $\bm{\tilde{X}}$ is the output from the preceding self-attention layer and $\bm{H}$ is the output of the last encoder block.
Within each encoder or decoder block, residual connections~\cite{He2016DeepRL} with layer normalisation~\cite{ba2016layer} are used across layers. As the last step of the decoder, an output projection layer is used to obtain $P(y_{i}|y_{0:i-1},\bm{X})$.


%% file: text/data_aug.tex
\presec
\section{Data Augmentation for DNC}
\postsec
\label{sec:data_aug}
\glsreset{daug}
Applications of \gls{dnc} are likely to be tasks with limited training data, such as speaker diarisation where each conversation or each meeting is treated as one input sequence. The issue of data scarcity is mitigated through the three proposed data augmentation techniques. They can also be combined. The data augmentation techniques have two, possibly competing, objectives. The first is to generate as many training sequences $\left(\bm{X},y_{1:N}\right)$ as possible. The second is for them to match the true data distribution $p\!\left(\bm{X},y_{1:N}\right)$ as closely as possible.
The exact setup of the proposed data augmentation techniques, their combination and application to speaker diarisation is in \sectdot{results}.

\presec
\subsection{Sub-Sequence Randomisation}
\postsec
\label{ssec:aug_subseq}
The first data augmentation technique is to simply train on multiple sub-sequences $\left(\bm{X}_{s:e},y_{s:e}\right)$ of the full sequence, where $s$ and $e$ are the random starting and ending 
indexes. While increasing the number of input sequences for \gls{dnc}, it also allows the same $\bm{x}_i$ to correspond to different $y_i$ in different sub-sequences since the specific ordering that cluster labels always start from `$\mathtt{1}$' and are incremented for any new cluster (\sectdot{neural_clustering}). Hence, \gls{dnc} is trained not to associate $\bm{x}_i$ with a fixed cluster label $y_i$.

\presec
\subsection{Input Vectors Randomisation}
\postsec
\label{ssec:aug_randomspk}
\begin{figure}[ht]
    \centering
    \includegraphics[width=\linewidth]{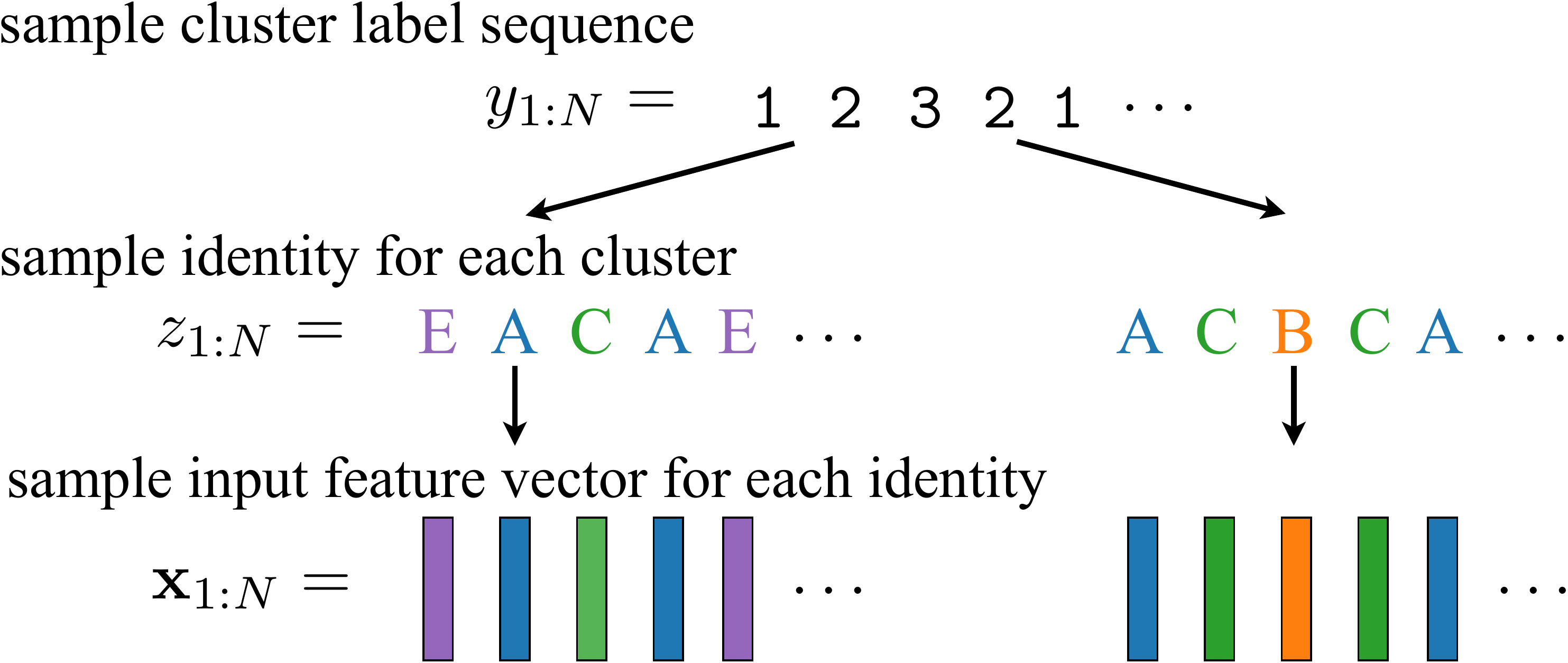}
    \caption{Examples of input vector randomisation generating two input sequences for one label sequence.}
    \label{fig:randomspk}
\end{figure}
The second technique, depicted in \figdot{randomspk}, modifies each training sequence by preserving its cluster label sequence $y_{1:N}$ such that every cluster label sequence is a true sample of the prior distribution $P(y_{1:N})$.
To achieve this goal, every cluster in each cluster label sequence is first reassigned to an identity randomly chosen from the training set, which results in a new identity sequence $z_{1:N}$. Next, for each $z_i$, a feature vector from the training set belonging to the identity $z_i$ is randomly chosen as $\bm{x}_i$.
In \ssectdot{data_aug_results}, two specific approaches to choosing  $z_i$ and $\bm{x}_i$ at the level of a meeting or the entire training set are given for speaker diarisation.

\presec
\subsection{Diaconis Augmentation}
\postsec
\label{ssec:aug_rotate}
\begin{figure}[ht]
    \centering
    \includegraphics[width=0.65\linewidth]{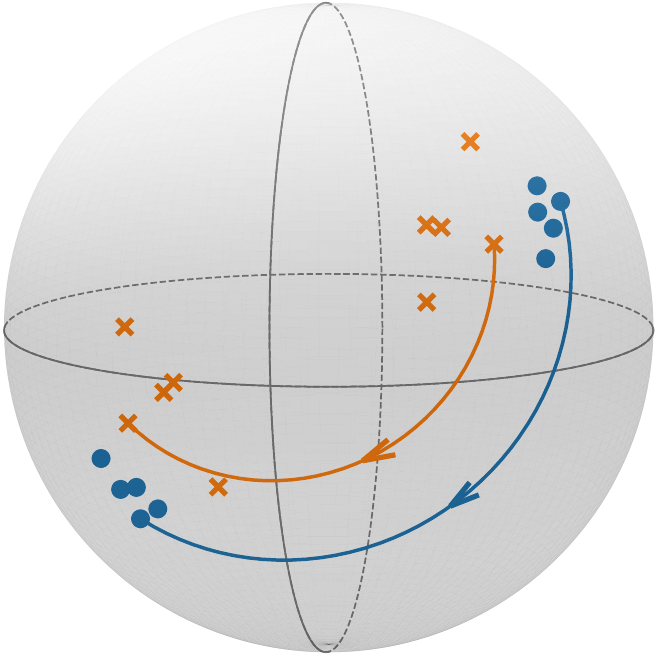}
    \caption{Diaconis Augmentation for two clusters. The rotated clusters form a new training example.}
    \label{fig:diaconis}
\end{figure}
The third technique, \gls{daug}, is applicable  when the feature vectors $\bm{x}_i$ are $L_2$-normalised, forming clusters on the surface of a hypersphere whose radius is the $L_2$-norm of the vectors. This is true for the speaker embeddings used in this paper. \gls{daug} rotates the entire input sequence $\bm{X}\in\mathbb{R}^{N\times D}$ to a different region of the hypersphere. This synthesises an entirely new training sequence pair $\left(\bm{X}',y_{1:N}\right)$ with previously unseen $\bm{x}_i'$. To do so, a random rotation matrix $\bm{R}\in\mathbb{R}^{D\times D}$ is sampled and $\bm{X}' = \bm{X}\bm{R}$. An example of such rotation is demonstrated in \figdot{diaconis}, with the paths of the rotation drawn. The algorithm for randomly sampling high dimensional rotation matrices was developed by Diaconis \etal~\cite{diaconisshahshahani_1987}, hence the name of our novel data augmentation algorithm. Diac-Aug can effectively prevent the model from overfitting to some particular regions on the hypersphere when data is limited.

%% file: text/setup.tex
\glsreset{sc}
\glsreset{ser}
\section{Diarisation Pipeline \& Clustering Setup}
\label{sec:setup}
In this paper, \gls{dnc} is applied to speaker clustering, the final stage of speaker diarisation. The speaker embeddings of the speech segments are clustered and each cluster represents a unique speaker. 
The speaker diarisation pipeline, clustering baseline and \gls{dnc} model setup are outlined below\protect\footnote{Our code is available at \href{https://github.com/FlorianKrey/DNC}{$\mathtt{https://github.com/FlorianKrey/DNC}$.}}. 
\presec
\subsection{Data and Segmentation}
\postsec
The \gls{mdm} data of the AMI meeting corpus \cite{carletta2005ami} is used for all experiments, where the official training (train), development (dev) and evaluation (eval) split is followed. The eight-channel audio data is first merged into a single stream using {\em BeamformIt} \cite{anguera2008beamf}.
\begin{table}[ht]
    \centering
    \begin{tabular}{c|ccc}
        \toprule
              & \#meetings & avg. duration & \#speakers  \\
        \midrule
        train & 147\footnote{Meetings $\mathtt{IS1003b}$ and $\mathtt{IS1007d}$ are removed as the \gls{mdm} audio is lost, after which 135 meetings are left. Since we limit the number of speakers to 4, each of the 3 five-speaker meetings is duplicated into 5 four-speaker meetings by removing one speaker at a time, which yields 147 meetings.}       &   37.9 min.    & 155 \\ 
        dev   & \;18     &   32.3 min.    & \;\;\;21\footnote{Among speakers in the dev set, 2 speakers appear in the training set.}\\
        eval  & \;16     &   34.0 min.    & \;16\\
        \bottomrule
    \end{tabular}
    \caption{Details of AMI corpus partitions used for both training the speaker embedding generator and training the \gls{dnc} model.}
    \label{tbl:dataset}
\end{table}

To compare the performance of spectral clustering and \gls{dnc}, perfect \acrlong{vad} is assumed by using the manual segmentation of the original dataset and stripping silences at both ends of each utterance. Some short segments are completely enclosed within longer segments, which are unrepresentative of the output generated by common segmentation schemes operating on single-stream audio. 
Therefore, such enclosed segments are removed from the training and test sets.

\presec
\subsection{Segment-level Speaker Embedding Generator}
\postsec
\label{ssec:dvec}
Clustering is performed based on speech segments, \ie one speaker embedding $\bm{x}_i$ per segment. $\bm{x}_i$ is obtained by averaging the window-level speaker embeddings within the corresponding speech segment, which are $L_2$-normalised both before and after averaging. 
The input window to the window-level embedding generator is around 2 seconds (215 frames, [-107,+106]) long. The embedding generator uses a \gls{tdnn}~\cite{peddintiTimeDelayNeural,kreyssig2018TDNN} with a total input context of [-7,+7], which is shifted from \{-100\} to \{+99\} (resulting in the overall input window of [-107,+106]). The output vectors of the \gls{tdnn} are combined using the self-attentive layer proposed in~\cite{sun2019diarselfatt}. This is followed by a linear projection down to the embedding size,
which is then the window-level embedding. The TDNN structure resembles the one used in the x-vector models \cite{snyder2018xvec} (\ie \gls{tdnn}-layers with the following input contexts: [-2,+2], followed by \{-2,0,+2\}, followed by \{-3,0,+3\}, followed by \{0\}). The first three \gls{tdnn}-layers have a size of 512, the third a size of 128 and the embedding size is 32.
The embedding generator is trained as a speaker classifier on the AMI training data with the \gls{asoft} loss~\cite{Liu2017sphereface} using HTK~\cite{youngHTK} and PyHTK~\cite{zhang2019pyhtk}. 
By using the \gls{asoft} loss combined with a linear activation function for the penultimate layer of the d-vector generator, the $L_2$-normalised window-level speaker embeddings,
and in turn the segment-level speaker embeddings, should be approximately uniformly distributed on the unit hypersphere.
Based on this assumption and the speaker embedding size being 32, the mean and variance of individual dimensions of speaker embeddings should be close to zero and ${1}/{32}$, respectively. Empirically, this assumption fits well for the mean and most dimensions for the variance. Variance normalisation for the \gls{dnc} models is done by scaling the embeddings by $\sqrt{32}$.


\presec
\subsection{Clustering}
\subsubsection{Spectral Clustering}
\postsec
\label{ssec:spectral_clustering}
The baseline uses the refined spectral clustering algorithm proposed in ~\cite{wang2018diarLSTM}, the input of which is the segment-level speaker embeddings described in \ssect{dvec}. Our implementation is based on the one published by~\cite{wang2018diarLSTM}, but the distance measure used in the K-means algorithm is modified from Euclidean to cosine similarity to align exactly with~\cite{wang2018diarLSTM}. The number of clusters allowed is set to be between two and four.

\presec
\subsubsection{Discriminative Neural Clustering Model}
\postsec
\label{ssec:dnc_model}
The Transformer used in the \gls{dnc} model contains 4 encoder blocks and 4 decoder blocks with dimension $D=256$. The total number of parameters is 7.3 million. The number of heads for the \gls{mha} is 4. The model architecture follows~\cite{Vaswani2017AttentionIA} and is implemented using ESPnet~\cite{watanabe2018espnet}. The Adam optimiser is used with a variable learning rate, which first ramps up linearly from 0 to 12 in the first 40,000 training updates and then decreases in proportion to the inverse square root of the number of training steps~\cite{Vaswani2017AttentionIA}. A dropout rate of 10\% is applied to all parameters. Considering that the input-to-output alignment for \gls{dnc} is strictly one-to-one and monotonic (see \ssectdot{dnc}, one cluster label needs to be assigned to each input vector), the source attention between encoder and decoder, represented as a square matrix, can be restricted to an identity matrix. For our experiments, the source attention matrix is masked to be a tri-diagonal matrix, \ie only the main diagonal and the first diagonals above and below are non-zero. We refer to this type of restricted source attention as \emph{diagonal local attention}. 
This restriction was found to be important for effective training of \gls{dnc} models.

\presec
\subsection{Evaluation}
\postsec
Diarisation systems are evaluated based on their \gls{der}, \ie the combined duration of missed speech, false alarm speech and speaker error over the duration of the audio. To evaluate the clustering algorithm with given manual segmentation, \gls{der} becomes \gls{ser}. The NIST scoring script is used with a 250 ms collar and overlapping-speech regions are excluded.

%% file: text/curlurn.tex
\glsreset{sc}
\presec
\section{Experimental Results}
\postsec
\label{sec:results}

Experiments were designed to evaluate the performance of \gls{dnc} models for diarisation in general and to evaluate the proposed data augmentation methods. Furthermore, the experiments also demonstrate the necessity for a curriculum learning scheme. Finally, \gls{dnc} and \gls{sc} are evaluated and compared in term of \gls{ser} for speaker diarisation.

\subsection{Data Augmentation}
\postsec
\label{ssec:data_aug_results}
Sub-sequence randomisation (\ssectdot{aug_subseq}) was applied to the training set for all experiments, in addition to other augmentation techniques.
For the dev set used as the validation set in training, this augmentation technique is applied exclusively.
All augmentation techniques are compared for sub-meetings of length 50 in \tbl{data_aug}. Per original meeting, which dictates the label sequence $y_{1:N}$, 5000 sub-meetings with 50 segments each were generated and augmented using the techniques from \sectdot{data_aug}. The full eval set meetings were split into sub-meetings with at most 50 segments.

For diarisation on the AMI corpus, two variants of the proposed input vector randomisation were compared. The first one, referred to as \globala randomisation, samples speakers (the set of possible identities in $z_{1:N}$) uniformly at random from the entire training set. Each feature vector $\bm{x}_i$ is then randomly chosen from all segment embeddings of the training set belonging to speaker $z_i$. The second variant, called \meetinga randomisation, first samples a meeting from the training set. Its speakers form the possibilities for $z_{1:N}$ and vectors $\bm{x}_i$ are sampled only from those vectors belonging to speaker $z_i$ that are in the sampled meeting. Thus, \meetinga randomisation preserves correlations among the speaker embeddings within one meeting. For \globala and \meetinga randomisation, all segments in the training set longer than 2 seconds were split into as many segments of at least 1 second as possible. This splitting operation is only used to obtain additional segment embeddings for sampling $\bm{x}_i$, leaving the label sequences untouched. 
\begin{table}[ht]
    \centering
    \begin{tabular}{l|cc}
        \toprule
         randomisation & w/o \gls{daug} & w/ \gls{daug}\\
        \midrule
         \nonea    & 20.19 & 15.25\\
         \globala & \bf{14.47} & 19.80\\
         \meetinga & 23.03 & \bf{13.57}\\
        \bottomrule
    \end{tabular}
    \caption{\%\glspl{ser} of different data augmentation techniques for sub-meetings with a length of 50 segments on the eval set. Input vector randomisation schemes, \nonea, \globala and \meetinga are combined with \gls{daug}.} 
    \label{tab:data_aug}
\end{table}

The \gls{ser} of a \gls{dnc} model trained on non-augmented data (\nonea in Table~\ref{tab:data_aug}), is 20.19\%. Using \globala augmentation reduces the \gls{ser} to 14.47\%, whilst \meetinga augmentation only achieves an \gls{ser} of 23.03\%. Neighbouring embeddings are challenging to cluster due to overlapping speech and similarities in the acoustic environment. For short sub-meetings, \meetinga randomisation can move such neighbouring embeddings into separate sub-meetings. Hence, for short sub-meetings \meetinga might generate not very typical data whilst providing far less augmentation than \globala.

The trend is different after applying \gls{daug} (\nth{2} column of \tbl{data_aug}). Using \nonea achieves an \gls{ser} of 15.25\%, and the result for \meetinga reduces to 13.57\%, which shows that \gls{daug} generates fairly typical data. \ssect{dvec} showed that the assumptions behind \gls{daug} are not perfect, which explains the performance drop of \globala. The number of speaker groups available from \globala (${\sim} C^{155}_4$) is larger than the number of generated sub-meetings. Thus, \gls{daug} does not increase the number of speaker groups seen by the \gls{dnc} model when used together with \globala.

\presec
\subsection{Curriculum Learning Scheduling}
\postsec
Training directly with full-length meetings, using \meetinga and \gls{daug} for augmentation, lead to a high \gls{ser}. A \gls{cl} \cite{elman1993learnsmall,Bengio2009CL} approach, \ie training in increasingly difficult stages, is used. The \gls{cl} approach used in this paper is to train with increasingly long sub-meetings as training examples as learning long sequences is hard.
First, sub-meetings with 50 segments were chosen to reduce the input space significantly. By containing fewer speakers on average, such sub-meetings represent an easier task. Subsequently, the maximum length of the sub-meetings increases to 200, then to 500 and then to the full length (up to 1682 segments). The model was trained to convergence in all four stages.
In each of the last three stages, each sub-meeting has a variable length of between 50\% and 100\% of the maximum length. Meetings of the eval set were split into as few sub-meetings as possible, each no longer than the maximum length.
After training to convergence with data augmentation, the \gls{dnc} models are fine-tuned using only sub-sequence randomisation for augmentation (akin to \nonea in \tbl{data_aug}).
\begin{table}[ht]
    \centering
    \begin{tabular}{c|cc|cc}
        \toprule
        \#segments in & \multicolumn{2}{c|}{dev} & \multicolumn{2}{c}{eval} \\
        sub-meetings & PT & FT & PT & FT\\
        \midrule
        \phantom{0}50 & 18.44 & 17.94 & 13.57 & 13.90\\
        200  & 23.82 & 20.51 & 16.92 & 16.75\\
        500  & 25.48 & 21.89 & 17.73 & 18.39\\
        \;\;all & 28.13 & 26.15 & 20.65 & 16.92\\
        \bottomrule
    \end{tabular}
    \caption{Results of \gls{dnc}'s \gls{ser} performance for the four \gls{cl} stages. Comparison between only pre-training (PT) on data augmented with \meetinga and \gls{daug} and finetuning (FT) afterwards on non-augmented data. The dev set was used as the validation set.}
    \label{tab:dnc_cl}
\end{table}

For the stages with a meeting lengths of above 50, for each original meeting, $10^4$ sub-meetings are generated using the augmentation techniques. When the maximum length is set to 200, applying the two best-performing techniques of \ssectdot{data_aug_results} results in an \gls{ser} of 19.14\% for \globala and 16.92\% for \meetinga combined with \gls{daug}. The later \gls{cl} stages use the latter combination of techniques for data augmentation.
Increasing the maximum sub-meeting length to 500 results in an \gls{ser} of 17.73\%. The \gls{ser} for full-length (34.0 minutes on average) meetings is 20.65\%. \tbl{dnc_cl} shows the results for \meetinga with \gls{daug} in column ``eval--PT''.

\begin{table}[ht]
    \vspace{0.5em}
    \centering
    \begin{tabular}{cr|c|c}
        \toprule
        \multicolumn{2}{c|}{sub-meeting length} & \multirow{2}{*}{\gls{dnc}} & \multirow{2}{*}{SC}\\
        \#segments & duration & &\\
        \midrule
        \phantom{0}50  &  2.8 mins & \bf{13.90} & 15.89\\
        200 &  9.7 mins & \textbf{16.75} & 22.38\\
        500 & 20.9 mins  & \textbf{18.39} & 23.56\\
        \;\;all & 34.0 mins & \textbf{16.92} & 23.95\\
        \bottomrule
    \end{tabular}
    \caption{Comparison of \gls{dnc} \vs \acrfull{sc} on the eval set for different meeting lengths. Fine-tuned \gls{dnc} models are used for comparison.}
    \vspace{-0.5em}
    \label{tab:dnc_sc}
\end{table}

\captionsetup[subfigure]{labelformat=empty}
\begin{figure*}[ht]
    \centering
    \begin{subfigure}[t]{0.24\textwidth}
        \centering
        \includegraphics[width=\linewidth]{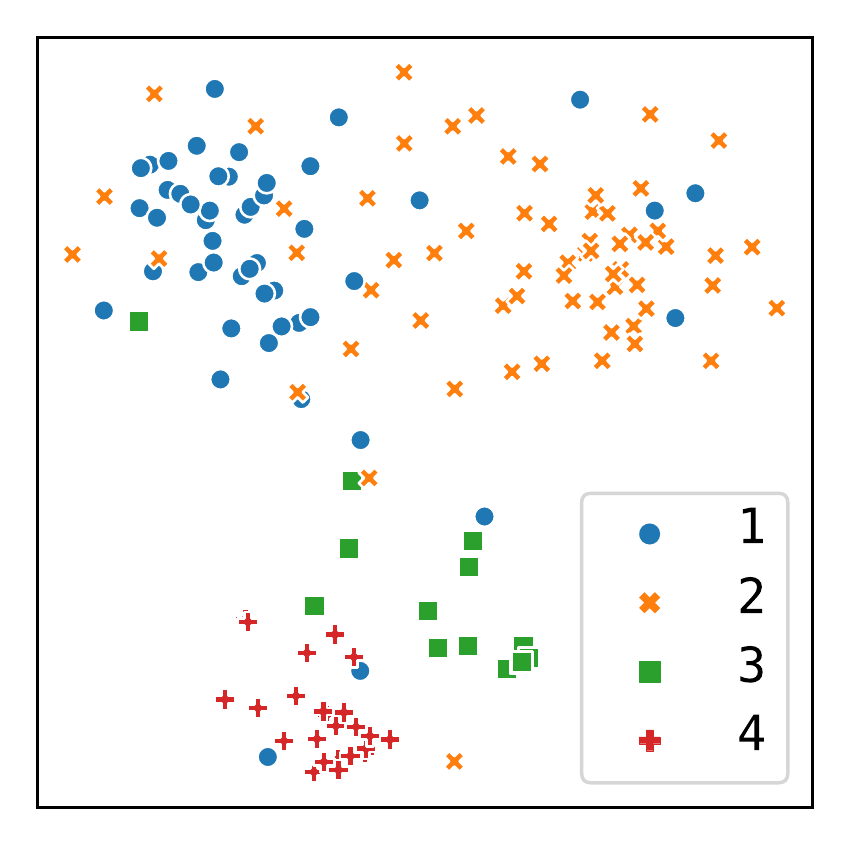}
        \caption{\normalsize (a) ground truth}
        \label{fig:ground_truth}
    \end{subfigure}
    \hfill
    \begin{subfigure}[t]{0.24\textwidth}
        \centering
        \includegraphics[width=\linewidth]{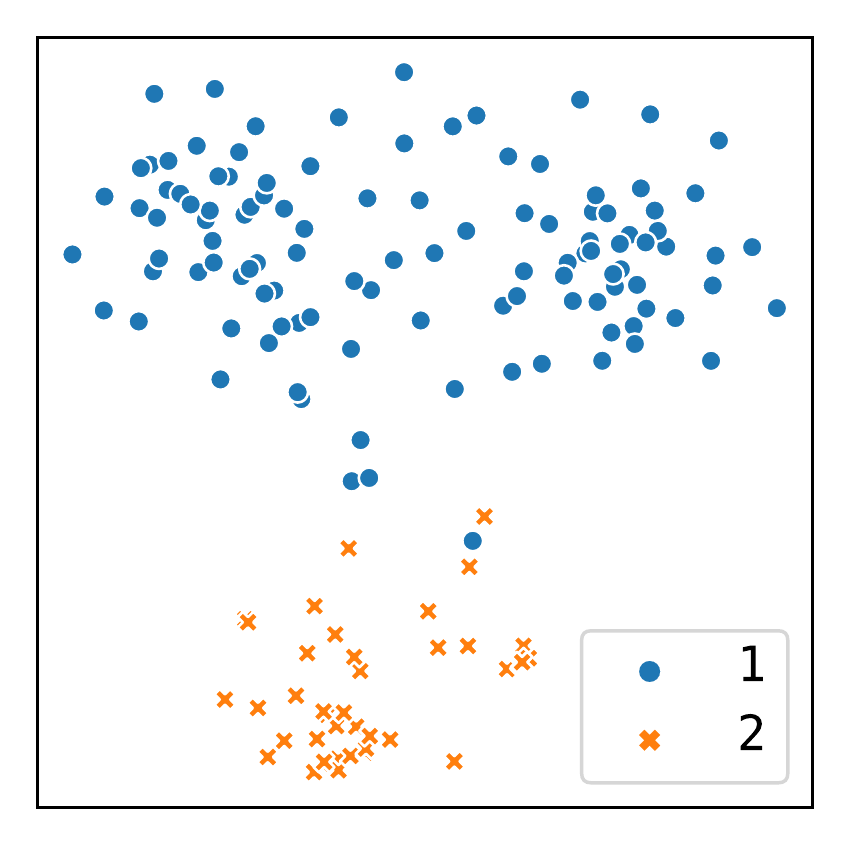}
        \caption{\normalsize (b) \gls{sc}}
        \label{fig:spectral_clustering}
    \end{subfigure}
    \hfill
    \begin{subfigure}[t]{0.24\textwidth}
        \centering
        \includegraphics[width=\linewidth]{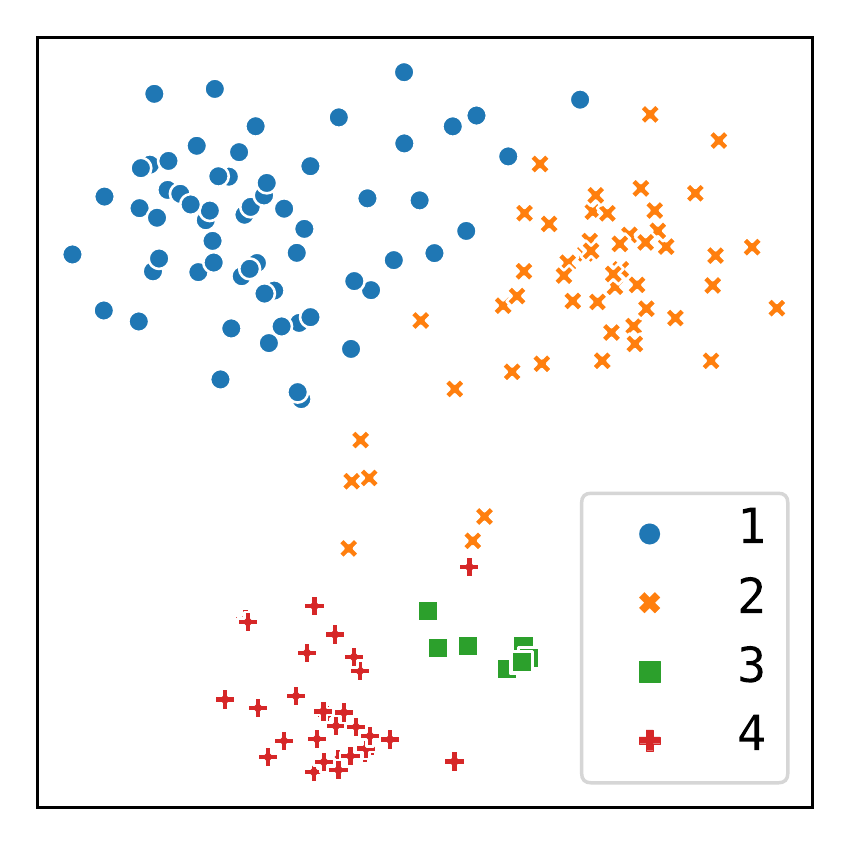}
        \caption{\normalsize (c) \acrshort{sc} known speaker number}
        \label{fig:spectral_clustering_oracle}
    \end{subfigure}
    \hfill
    \begin{subfigure}[t]{0.24\textwidth}
        \centering
        \includegraphics[width=\linewidth]{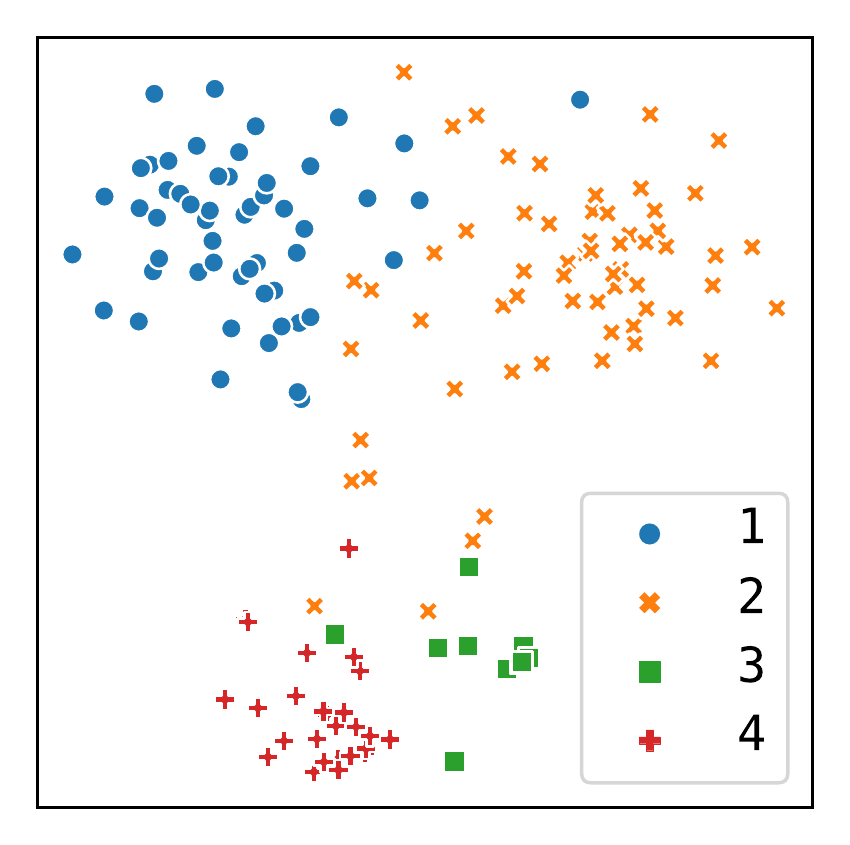}
        \caption{\normalsize (d) \gls{dnc}}
        \label{fig:dnc}
    \end{subfigure}
    \caption{Clustering results of different algorithms for one meeting. The 32D speaker embeddings are projected to 2D using t-SNE. In this example, (b) shows that \glsfirst{sc} fails to identify the correct number of speakers. Even when the correct number of clusters is given to spectral clustering as in (c), \gls{dnc} shows a better clustering result in (d).}
    \label{fig:clustering}
\end{figure*} 

\tbl{dnc_cl} also shows the results of finetuning the \gls{dnc} models of column ``eval--PT'' by training on data that uses only sub-sequence randomisation, but neither \meetinga nor \gls{daug} (see column ``eval--FT'').
While the eval set \gls{ser} increases after finetuning for some cases, the \gls{ser} on the dev set (our validation set) reduces in all cases. Therefore, the finetuned models are compared with spectral clustering in \tbl{dnc_sc} (column ``eval--FT'' in \tbl{dnc_cl} is column ``DNC'' in \tbl{dnc_sc}). The parameters of the spectral clustering algorithm are chosen to optimise the \gls{ser} on the dev set.
For all meeting lengths, the \gls{dnc} model outperforms spectral clustering. The finetuned \gls{dnc} model for the full meeting length achieves an \gls{ser} of 16.92\%, which is a reduction of 29.4\% relative to spectral clustering. The \gls{ser} of a \gls{dnc} model trained with \meetinga and \gls{daug}, but without \gls{cl}, is only 34.48\% after finetuning.

%% file: text/conclusion.tex
\presec
\section{Analysis}
\postsec
\label{sec:analysis}
\fig{clustering} visualises clustering results of a sub-meeting of the eval set using a 2D \acrshort{tsne}~\cite{maaten2008visualizing} projection of the 32D segment embeddings. In \figdot{ground_truth}, depicting the ground truth cluster labels, there are no clear cluster boundaries. Spectral clustering is only able to pick out two speakers in \figdot{spectral_clustering}, resulting in a high \gls{ser}. The \gls{dnc} model correctly recognises the existence of four speakers. \fig{spectral_clustering_oracle} and \figdot{dnc} show the comparison between the behaviour of spectral clustering (assuming the number of clusters is known) and the \gls{dnc} model. The cluster boundaries are relatively clear for spectral clustering. However, the \gls{dnc} approach has more complex boundaries and shows multiple points of cross over to other clusters, similar to \figdot{ground_truth}.
Especially, when two clusters have significant overlap, spectral clustering seems to make more errors, \eg it wrongly assigns many samples from speaker 2 to speaker 1. By comparison, the \gls{dnc} model can split these two confusing clusters better. In this case, \gls{dnc} yields a much lower \gls{ser} than spectral clustering. 

\presec
\section{Conclusions}
\postsec
\glsreset{dnc}
\label{sec:conclusion}


We proposed \gls{dnc}, a novel, supervised clustering approach based on sequence-to-sequence neural networks. \gls{dnc} models in the form of the Transformer model were trained using three effective, targeted data augmentation techniques and curriculum learning. The data augmentation techniques are sub-sequence randomisation, input vector randomisation, and Diaconis augmentation. It was shown that these techniques were themselves effective but also can be combined for even better performance.
On the challenging AMI MDM meeting data with up to 4 speakers in each meeting, 
\gls{dnc} was shown to perform much better for speaker clustering than a strong spectral clustering baseline.

Given that the neural network used to extract the d-vectors is differentiable, \gls{dnc} can be extended to an end-to-end trainable model for speaker diarisation. Furthermore, instead of optimising the segment level cross-entropy, \gls{dnc} could be trained to directly optimise the \gls{der} based on a minimum Bayesian risk loss~\cite{Prabhavalkar2018MinimumWE} or the REINFORCE algorithm~\cite{Williams1992SimpleSG}. To perform online diarisation, the Transformer-based \gls{dnc} models could be modified to use forms of monotonic attention~\cite{Arivazhagan2019MonotonicIL,ma2019monotonic}.
